\documentstyle[12pt,times]{article}
\newlength{\abstractwidth}
\renewcommand{\thefootnote}{\fnsymbol{footnote}}
\renewcommand{\thanks}[1]{\footnote{#1}} 
\newcommand{\starttext}{
\setcounter{footnote}{0}
\renewcommand{\thefootnote}{\arabic{footnote}}}

\newcommand{\be}{\begin{equation}}
\newcommand{\bea}{\begin{eqnarray}}
\newcommand{\eea}{\end{eqnarray}}
\newcommand{\beq}{\begin{equation}}
\newcommand{\ee}{\end{equation}}
\newcommand{\eeq}{\end{equation}}
\newcommand{\N}{{\cal N}}

\newcommand{\dmup}{\partial_{\mu'}}
\newcommand{\dnup}{\partial_{\nu'}}
\newcommand{\dmu}{\partial_\mu}
\newcommand{\dnu}{\partial_\nu}

\def\ba{\begin{eqnarray}}
\def\ea{\end{eqnarray}}

\begin{document}
\begin{titlepage}
\bigskip
\hskip 3.7in\vbox{\baselineskip12pt
\hbox{MIT-CTP-2824}
\hbox{UCLA/99/TEP/1}
\hbox{hep-th/9902042}}
\bigskip\bigskip\bigskip\bigskip

\centerline{\Large \bf Graviton and gauge boson propagators in 
\protect \boldmath $AdS_{d+1}$}

\bigskip\bigskip
\bigskip\bigskip

\centerline{ Eric D'Hoker$^{a}$, Daniel Z. Freedman$^{b,c}$, 
Samir D. Mathur$^{b}$, }
\medskip
\centerline{Alec Matusis$^{b}$  
and Leonardo Rastelli$^{b,}$\footnote[1]{\tt dhoker@physics.ucla.edu, dzf@math.mit.edu, me@ctpdown.mit.edu, alec\_m@ctp.mit.edu, rastelli@ctp.mit.edu.}}
\bigskip
\bigskip
\centerline{$^a$ \it Department of Physics}
\centerline{ \it University of California, Los Angeles, CA 90095}
\bigskip
\centerline{$^b$ \it Center for Theoretical Physics}
\centerline{ \it Massachusetts Institute of Technology}
\centerline{ \it Cambridge, {\rm MA}  02139}
\bigskip
\centerline{$^c$ \it Department of Mathematics}
\centerline{ \it Massachusetts Institute of Technology}
\centerline{\it Cambridge, {\rm MA} 02139}
\bigskip\bigskip

\begin{abstract}

We construct the gauge field and  graviton propagators in Euclidean 
$AdS_{d+1}$
space-time by two different methods. In the first method the gauge invariant
Maxwell or linearized Ricci operator is applied directly to bitensor ansatze
for the propagators which reflect their gauge structure. This leads to a
rapid determination of the physical part of the propagators in terms of
elementary functions. The second method is a more traditional approach using
covariant gauge fixing which leads to a solution for both physical
and gauge parts of the propagators. The gauge invariant parts agree 
in both methods.
\end{abstract}

\end{titlepage} 
\starttext
\baselineskip=18pt
\setcounter{footnote}{0}

\section{Introduction}

One important aspect of the AdS/CFT correspondence
conjectured in \cite{maldacena,polyakov,witten} is the calculation of correlation functions in Type IIB
supergravity on $AdS_5 \times S^5$ \cite{vannieuv} in order to study the large N
strong coupling dynamics of
$\N=4$ superconformal Yang-Mills theory.

Bulk to bulk Green's functions, describing propagation between two interior 
points of $AdS$, are required to compute  4-point functions
\cite{strings, liutseytlin, august, dhfgauge, liu, dhfscalar}.
Scalar and spinor propagators for general mass have been discussed in
the old literature \cite{prop} on $AdS$ field theory. The Feynman gauge propagator
for massless gauge bosons, was obtained in \cite{allenjacobs}. The result involves
transcendental functions which cancel for physical components of the gauge
field. This motivated a recent simpler derivation \cite{dhfgauge} which used a 
convenient  bitensor ansatz to obtain the physical part of the propagator
as
a simple  algebraic function of the chordal distance variable. There are older
covariant treatments of the graviton propagator \cite{allenturyn,
turyn, mottola} in
de Sitter space (of constant positive curvature) which again involve
transcendental functions, and there is recent work on the gauge boson and
graviton propagators for $AdS$ in non--covariant  gauges \cite{liutseytlin}.

In the present paper, we shall construct the gauge boson and graviton 
propagator in $AdS_{d+1}$  using two different methods, both covariant
under the isometry group $SO(d+1,1)$ of the Euclidean AdS space. 

The first
method is a modification of that of \cite{dhfgauge}. The key step 
 is a pair of ansatze for the bitensor propagators 
which naturally
separate gauge invariant parts from gauge artifacts. The latter do not
contribute for conserved sources of the fields. The gauge invariant
components can then be rather simply obtained by substituting the ansatz
in the Maxwell or linearized Ricci equations. They are given by
elementary functions. Gauge fixing is unnecessary because we work on the
subspace of conserved sources. 
In the second method the propagators are
constructed in the Landau gauge. This method is closely related to the work
of \cite{allenjacobs, allenturyn, turyn, mottola}, with some new
technical simplifications. 
In particular we find a simple and general way to solve the
inhomogeneous equation that arises for the transverse traceless part
of the propagator.
It is
shown explicitly that the gauge invariant parts of the propagators agree in
both methods.

These results have immediate applications to $4$-point functions in
the $AdS/CFT$ correspondence. Only the gauge invariant part of the
propagators is required, since they are coupled to conserved currents
or stress tensors. The amplitude for gauge boson exchange between external
scalar fields was presented in \cite{dhfgauge}. The calculation of graviton exchange
for even $d$ and general integer scaling dimension $\Delta$ of external
scalars is complete and will be presented shortly \cite{amplitude}.

The paper is organized as follows. In Section 2 we introduce standard
notations for Euclidean $AdS_{d+1}$ and apply the new method to obtain
the gauge boson propagator. This method is then applied to the graviton
propagator in Section 3. The more traditional Landau gauge methods are
applied to the gauge
and graviton propagators in Sections 4 and 5, respectively. 

\section{ \bf  New method for the gauge propagator}
\setcounter{equation}{0}
We work on the Euclidean continuation of $AdS_{d+1}$, viewed
as the upper half space, $z_0 >0$, in ${\bf R}^{d+1}$ with
metric $g_{\mu
\nu}$
 given by
\be
  ds^2 = \sum _{\mu, \nu=0} ^ d g_{\mu \nu} dz_\mu dz_\nu
       ={1 \over z_0^2} (dz^2_0 + \sum ^d_{i=1} \, dz^2_i)
\, .
\label{2.1}
\ee The $AdS_{d+1}$ scale has been set to unity, and the
metric above has constant negative curvature $R=-d(d+1)$. It
is well known that invariant functions and tensors on
$AdS_{d+1}$ are most  simply expressed in terms of the
chordal distance variable  $u$, defined by
\be
  u \equiv
  {(z-w)^2 \over 2z_0 w_0}
\label{2.2}
\ee
where $(z-w)^2 = \delta_{\mu \nu}(z-w)_{\mu} (z-w)_{\nu}$
is the ``flat Euclidean distance".

\medskip

The action of an abelian gauge field coupled to a conserved
current source $J^\mu$ in the $AdS_{d+1}$ background is
\be
  S_A = \int d^{d+1}z \sqrt g \big [ {1 \over 4} F^{\mu \nu}
F_{\mu \nu} -
      A_{\mu} J^{\mu} \big ] \, .
\label{2.3}
\ee We have not included a gauge fixing term, because we work
directly on the restricted space of conserved currents and
look for a  solution of the Euler-Lagrange equation in the
form
\be A_\mu(z) = \int d^{d+1}w  \sqrt g G_{\mu \nu'}(z,w)
J^{\nu'}(w)\
\label{2.4}
\ee with bitensor propagator $G_{\mu \nu'} (z,w)$. We see
that the  propagator must satisfy the AdS-covariant equation
$(\partial_{\mu} = {\partial \over \partial z^{\mu}}$ and
$\partial_{\mu '} = {\partial \over \partial w^{\mu '}})$
\be
  D^{\mu} \partial_{[\mu} G_{\nu ] \nu'}  = -
  g_{\nu \nu'}  \delta (z,w) + \partial_{\nu'} \Lambda _\nu
(z,w)\,
\label{2.5}
\ee Here, $\Lambda_\nu$ is a vector function, which acts as a
pure gauge term, and will cancel out when integrating (\ref{2.5})
against any conserved current $J^{\nu '}$.
\medskip

Any bitensor can be expressed \cite{allenjacobs} as a sum of two linearly
independent forms with scalar coefficients.  Following 
\cite{dhfgauge} we  choose two independent  bitensors derived from
the  biscalar variable $u$, namely
\be
  \partial_{\mu} \partial_{\nu'}u= -
  {1 \over z_0w_0} \big [ \delta_{\mu \nu'} +
  {1 \over w_0} (z-w)_{\mu} \delta_{\nu'0} +
  {1 \over z_0} (w-z)_{\nu'}\delta_{\mu 0} -
  u \delta_{\mu 0} \delta_{\nu'0} ]
\label{2.6}
\ee and $\partial_{\mu}u \partial_{\nu'}u$ with
\bea\label{pippo}
  \partial_{\mu} u &=& {1 \over z_0}
     [(z-w)_{\mu} / w_0 - u \delta_{\mu 0}] \\
   \partial_{\nu'} u &=& {1 \over w_0}
     [(w-z)_{\nu'} / z_0 - u \delta_{\nu' 0}] \, .\nonumber
\eea
\medskip The propagator can be represented as a
superposition of  the bitensors $\partial_\mu \partial_{\nu'}
u$ and
$\partial_{\mu} u\, \partial_{\nu'} u$  times scalar functions of
$u$. It is more convenient to use the equivalent form \cite{dhfgauge}
\be
  G_{\mu \nu'}(z,w) =- (\partial_{\mu} \partial_{\nu'}u)
  F(u) + \partial_{\mu} \partial_{\nu'} S(u)
\label{2.8}
\ee in which $F(u)$ describes the propagation of the physical
components of
$A_\mu$ and $S(u)$ is a gauge artifact. We see that $S(u)$
drops out of the invariant equation (\ref{2.5}) and the integral
solution (\ref{2.4}) and can be discarded. Thus we must determine
only $F(u)$ which we will find by substitution of (\ref{2.8}) into
(\ref{2.5}).

\medskip

For this we need certain properties of derivatives of $u$,
most of which were derived in \cite{dhfgauge} and which we list here.
(Some of these will be needed only later.)
\bea
&&   \Box u= D^{\mu} \partial_{\mu} u = (d+1)(1+u)  
\label{2.9a} \\
&&      D^{\mu} u \ \partial_{\mu} u = u (2+u)   \label{2.9b} \\
&&      D_{\mu} \partial_{\nu} u = g_{\mu \nu}(1+u)   \label{2.9c}
\\
&&      (D^{\mu}u)\ (D_{\mu} \partial_{\nu}\partial_{\nu'}u)
          = \partial_{\nu}u \partial_{\nu'}u   \label{2.9d} \\
&&      (D^\mu u) \ (\dmu \dnup u)  = (1+u) \dnup u \label{2.9e}
\\
&&      (D^\mu \dmup u) \ (\dmu \dnup u)  = g_{\mu ' \nu '} +
\dmup u \dnup u  \label{2.9f} \\
&&      D_\mu \dnu \dnup u  = g_{\mu \nu} \dnup u  \label{2.9g}
\eea
 With the help of (\ref{2.9a}-\ref{2.9g}), we readily find
\bea\label{paperino}
  D^{\mu} \partial _{[\mu} G_{\nu ] \nu '}  &=& \dnu \dnup u
\big [ -u(2+u) F'' (u)  -d (1+u) F' (u) \big ] \\ &&+
\dnu u \dnup u \big [ (1+u) F'' (u) + d F' (u) \big ] \,
.
\nonumber\eea
For distinct $z$ and $w$, the $\delta$-function term in
(\ref{2.5}) may be ignored. Using $AdS$-invariance, we also have
$\Lambda _\nu = \dnu u \ \Lambda (u)$ in (\ref{2.5}). Using (\ref{paperino})
we find that (\ref{2.5}) gives the set of two second order
differential equations for the functions $F(u)$ and $\Lambda
(u)$, namely
\bea
u(2+u) F'' + d(1+u) F' & = - \Lambda & \label{2.11a}
\\ (1+u) F'' + d F' & = \Lambda ' \, . & \label{2.11b}
\eea
Now, (\ref{2.11b}) is readily integrated to give $(1+u) F' +
(d-1) F = \Lambda$. (An integration constant is chosen to
vanish so that F(u) vanishes on the boundary.) Eliminating
$\Lambda$ between (\ref{2.11a}) and (\ref{2.11b}) gives
\be u(2+u) F'' + (d+1) (1+u) F' +(d-1)F =0\, ,
 \label{2.12}
\ee which agrees with \cite{dhfgauge} and is just the invariant
equation
\be (\Box - m^2)F(u)=0
\label{2.13}
\ee
for the propagator of a scalar field of mass $m^2= -(d-1)$.
The solution is the elementary function, normalized
to include the $\delta$  function in (\ref{2.5}),
\be F(u) = {\Gamma \left ( {d-1 \over 2} \right ) \over ( 4
\pi) ^{(d+1)/2}}
\big [ u(2+u) \big ] ^{-(d-1)/2} \, .
\label{2.14}
\ee As a further check on the consistency of this approach,
we note that $\Lambda$ vanishes as $1/u^{d+1}$ on the
boundary. The current $J^{\nu'}(w)$ is also expected to
vanish, and there is then no surface term which could negate
the statement following (\ref{2.5}) that the gauge term cancels
out.

\medskip
There is an even faster variation of this method, a bit less clear
pedagogically, which we now outline. The advantage is that we need not deal
with the gauge function $\Lambda$ nor the coupled equations (\ref{2.11a}-\ref{2.11b}).
Instead we insert the ansatz (\ref{2.8}), with $S(u)$ dropped, in the integral (\ref{2.4}) and
apply the Maxwell operator inside the integral. After some calculation
using (\ref{2.9a}-\ref{2.9g}) we find
\bea
D^\mu \partial_{[\mu} A_{\nu]}(z)&=&-\int\,d^{d+1}w\sqrt{g}\;\bigg[
\dnu\dnup u \Box F - d \dnup u \dnu F - \dmu\dnup u D^\mu \dnu F
\bigg] J^{\nu'}(w)\\&=&-\int\,d^{d+1}w\sqrt{g}\;\bigg[\dnu\dnup u
(\Box F +(d-1) F) - \dnu\dnup ((d-1) F + (1+u) F')\bigg]\nonumber\,.
\eea
The manipulations of derivatives which lead from the first to second line
are designed so that the last term is a total derivative with respect to
$w$. This can be partially integrated and dropped by current 
conservation. One then finds that
\be
D^\mu \partial_{[\mu} A_{\nu]}(z)=-\int\,d^{d+1}w\sqrt{g}\;\dnu\dnup u
(\Box F +(d-1) F)J^{\nu'}(w)\;=\;-J_\nu(z)
\ee
must hold for all choices of current source $J_\nu(z)$. Therefore $F(u)$
must satisfy (\ref{2.13}) with $-\delta(z,w)$ on the right side, and the solution with
fastest decay on the boundary is (\ref{2.14}) as before.

The method of this section is somewhat simpler than that of \cite{dhfgauge} in which
a version of (\ref{2.5}) with gauge-fixing term added to the wave operator and
$\Lambda_\nu=0$ was solved. The immediate extension of that method to the
graviton propagator does not work readily because it leads to intricately
coupled equations for 5 scalar functions. The new method is far simpler,
as we will now see. 


\section{New method for the graviton propagator}
\setcounter{equation}{0}
The gravitational action in $d+1$ dimensional Euclidean space
with cosmological constant $\Lambda$  and covariant matter
action ($S_m$) is
\be S_g = {1\over2\kappa^2} \int d^{d+1} z \sqrt g (R - \Lambda) +
S_m\ .
\label{3.1}
\ee Henceforth we set $\kappa=1$. The first variation with
respect to
$g^{\mu \nu}$ gives the Euler-Lagrange equation
\be R_{\mu \nu} - {1\over 2} g_{\mu \nu} (R-\Lambda)  = T_{\mu \nu} \ .
\label{3.2}
\ee
We take $\Lambda=-d(d-1)$ so that without sources ($T_{\mu
\nu}=0$), we obtain Euclidean AdS space with $R=-d(d+1)$ as
the maximally symmetric solution of (\ref{3.2}).

\medskip A matter source produces a fluctuation $h_{\mu
\nu}=\delta g_{\mu\nu}$ about the AdS metric $g_{\mu \nu}$.
To obtain the graviton propagator it is sufficient to
consider (\ref{3.2}) to linear order in $h_{\mu\nu}$. It is simpler
to work with the equivalent ``Ricci form'' of this equation,
namely
\be R_{\mu\nu} + d g_{\mu\nu} = \tilde T _{\mu \nu}
           \equiv T_{\mu \nu} - {1 \over d-1} g_{\mu \nu}
T_\sigma{}^\sigma \ .
\label{3.3}
\ee It is straightforward but tedious to use textbook results
\cite{landau} on the linearized Ricci tensor and commute
covariant derivatives to obtain the linearized equation
\be -  D^\sigma D_\sigma h_{\mu \nu} -  D_\mu D_\nu h_\sigma
{}^\sigma +  D_\mu D^\sigma h_{ \sigma \nu} +  D_\nu
D^\sigma h_{\mu \sigma} -2(h_{\mu \nu} - g_{\mu \nu}
h_\sigma {}^\sigma) = 2 \tilde T_{\mu \nu}
\label{3.4}
\ee
All covariant derivatives and contractions in (\ref{3.4}) are
those of the $AdS$ metric $g_{\mu\nu}$ of (\ref{2.1}). At this
point we could impose a gauge condition such as the deDonder
gauge, $D^\sigma h_{\sigma\mu}= \dmu h^\sigma_\sigma/2$,
which simplifies (\ref{3.4}), and an approach of this type is
pursued in Sec 4. Here, however, we will generalize the
method used for the gauge propagator in Section 2 and work more
directly with (\ref{3.4}) which is invariant under infinitesimal
diffeomorphisms, $\delta h_{\mu\nu}= D_\mu V_\nu + D_\nu
V_\mu$.
\medskip We represent the solution of (\ref{3.4}) or,
equivalently, the linearization of (\ref{3.2}), as the integral
\be h_{\mu\nu}(z)=\int d^{d+1}w \sqrt g G_{\mu \nu;\mu ' \nu
'}(z,w)
  T^{\mu'\nu'}(w)
\label{3.5}
\ee This defines the propagator which must then satisfy
\bea \label{3.6}
 &-&  D^\sigma D_\sigma G_{\mu \nu;\mu ' \nu '}
  -  D_\mu D_\nu G_{\sigma} {}^\sigma {}_{\mu ' \nu '}
  +  D_\mu D^\sigma G_{ \sigma \nu; \mu ' \nu '} \\
  &+&  D_\nu D^\sigma G_{\mu \sigma ; \mu ' \nu '}
-2(G_{\mu \nu; \mu ' \nu '} - g_{\mu \nu} G_\sigma {}^\sigma
{}_{;\mu ' \nu'}) \nonumber\\  
 &=&
 \Bigl(g_{\mu \mu'}g_{\nu \nu'} +g_{\mu \nu'}g_{\nu \mu'} -  {2\over d-1}
g_{\mu\nu}g_{\mu' \nu'}\Bigr)
\delta(z,w)  + D_{\mu '} \Lambda _{\mu \nu;\nu'} + D_{\nu '}\Lambda _{\mu \nu;\mu'}
\nonumber\eea
where $\Lambda_{\mu\nu,\nu'}(z,w)$ represents a
diffeomorphism  (in the ${}'$ coordinates) whose contribution
will vanish when (\ref{3.6}) is integrated against the covariantly
conserved stress tensor $T^{\mu'\nu'}$.

Our strategy now is to solve (\ref{3.6}) by expanding  $G_{\mu
\nu;\mu ' \nu '}$ and  $\Lambda _{\mu \nu;\mu'}$ in an
appropriate basis of bitensors. One basis of fourth rank
bitensors obtained by consideration of the geodesic between
the points $z$ and $w$ was proposed by \cite{allenturyn} and will be
used in the next section. We start with a very similar basis
of 5 independent  fourth rank bitensors constructed from the
basic forms $\dmu\dmup u, \dmu u$, and
$\dmup u$ defined in (\ref{2.6}-\ref{pippo})
\bea\label{3.7}
 T^{(1)} _{\mu \nu;\mu' \nu'} &=&
 g_{\mu \nu} \ g_{\mu ' \nu'}
 \\ T^{(2)} _{\mu \nu;\mu' \nu'} & = &
 \dmu u \ \dnu u \ \dmup u \ \dnup u
\nonumber \\ T^{(3)} _{\mu \nu;\mu' \nu'} & = &
 \dmu \dmup u \ \dnu \dnup u + \dmu \dnup u \ \dnu \dmup u
 \nonumber\\ T^{(4)} _{\mu \nu;\mu' \nu'} & = &
 g_{\mu \nu} \ \dmup u \ \dnup u + g_{\mu ' \nu'} \ \dmu u \
\dnu u
 \nonumber\\ T^{(5)} _{\mu \nu;\mu' \nu'} & = &
 \dmu \dmup u \ \dnu u \dnup u + \dnu \dmup u \ \dmu u \dnup
u
 + (\mu' \leftrightarrow \nu')
\nonumber\eea
and a general ansatz for the propagator as the
superposition
\be
G_{\mu\nu;\mu'\nu'} = \sum_{i=1}^{5} T_{\mu\nu;\mu'\nu'}^{(i)}
A_{(i)}(u)
\label{3.8}
\ee
However, it is very advantageous to reorganize things so
as to separate physical components of the propagator from
gauge artifacts, and we therefore postulate the form
\bea \label{3.9}
G_{\mu \nu ;\mu' \nu'}& =& (\partial_\mu \partial_{\mu'} u\,\partial_\nu \partial_{\nu'} u+\partial_\mu \partial_{\nu'} u\,\partial_\nu \partial_{\mu'} u) \,G(u) +g_{\mu \nu} g_{\mu' \nu'} \,H(u) \\
&& +\,\partial_{(\mu} [\partial_{\nu)} \partial_{\mu'} u \,\partial_{\nu'}u \,X(u)]+ \partial_{(\mu'} [\partial_{\nu')} \partial_{\mu} u \,
\partial_{\nu}u \,X(u)] \nonumber \\
&&  +\,\partial_{(\mu} [\partial_{\nu)}u \,\partial_{\mu'} u \,\partial_{\nu'}u \,Y(u)] +\partial_{(\mu'} [\partial_{\nu')}u \,\partial_{\mu} u \,\partial_{\nu}u \,Y(u)] \nonumber \\
&& + \,\partial_\mu [\partial_\nu u\,Z(u)] \,g_{\mu' \nu'} + \partial_{\mu'} [\partial_{\nu'}u \,Z(u)] \,g_{\mu \nu} \,.\nonumber
\eea
where $(\cdot)$ denotes symmetrization with strength 1.
The terms involving $X$, $Y$ and $Z$ are  gradients with respect
to $z$ or $w$. The gradients with respect to $z$ are diffeomorphisms
which are annihilated in (\ref{3.6}) and 
give an irrelevant modification of $h_{\mu \nu}$ in (\ref{3.5}). The gradients with respect to $w$   
contribute to the left side of (\ref{3.6}) as
$D_{\mu'}\dnup(\cdots)$ and can be  absorbed by changing
$\Lambda_{\mu\nu;\nu'}$ on the right side. 
Thus we can restrict
our  attention to the physical $G(u)$ and $H(u)$ terms
provided we are sure that the form (\ref{3.9}) is actually
equivalent to (\ref{3.8}). It is not hard to convince oneself that
this is the case by applying derivatives in (\ref{3.9}) to obtain
a set of five equations relating each $A_{(i)}(u)$ to a
combination of
$G, H$, and derivatives of $X$, $Y$ and $Z$. These expressions
are simple enough that one can readily deduce that the
$A_{(i)}$ are uniquely determined by $G,H,X,Y,Z$ and vice
versa. There remains to write an analogous tensor
decomposition of the diffeomorphism term on right side of
(\ref{3.6}) and we use the following
\bea 
\label{3.10}
\Lambda _{\mu \nu; \nu'}& = & ~ g_{\mu \nu} \dnup u A(u)
 + \dmu u \dnu u \dnup u C(u)\\ & + & (\dmu \dnup u \dnu u +
\dnu \dnup u \dmu u) B(u)
\nonumber\eea

The stage is now set for the next part of our work
which is to substitute the $G$ and $H$ terms of (\ref{3.9}) and
the decomposition (\ref{3.10}) in (\ref{3.6}), apply all derivatives
using the relations (\ref{2.9a}-\ref{2.9g}), and express the result  as a
superposition of independent tensor terms. Actually 6
independent tensors appear because (\ref{3.6}) does not have the
symmetry under exchange of $z$ and $w$ that the propagator
itself possesses. The full calculation is important but
tedious, and we quote only the final result, which is
\bea
\label{3.11}
 & - &  D^\sigma D_\sigma G_{\mu \nu;\mu ' \nu '}
  -  D_\mu D_\nu G_{\sigma} {}^\sigma {}_{\mu ' \nu '}
  +  D_\mu D^\sigma G_{ \sigma \nu; \mu ' \nu '}
  +  D_\nu D^\sigma G_{\mu \sigma ; \mu ' \nu '}
\\  & - &2(G_{\mu \nu; \mu ' \nu '} - g_{\mu \nu}
G_\sigma {}^\sigma {}_{;\mu ' \nu'}) -D_{\mu'} \Lambda _{\mu
\nu;\nu'} - D_{\nu'} \Lambda _{\mu \nu; \mu'}
\nonumber\\  & = &
\ \ T^{(1)} \bigl [-u(2+u)H'' -2d (1+u) H' +2dH +4G
-2(1+u)G' -2(1+u)A \bigr ]
\nonumber\\ & + &
 T^{(2)} \bigl [ - 2 G'' - 2 C' \bigr ]  +
 T^{(3)} \bigl [-u(2+u) G'' - (d-1) (1+u) G' + 2(d-1) G -2B
\bigr ]
\nonumber\\ & + & g_{\mu \nu} \dmup u \dnup u \bigl [2(1+u) G'
+ 4(d+1) G -2A' \bigr ]
\nonumber\\ & + & g_{\mu ' \nu'} \dmu u \dnu u \bigl [-2 G'' -
(d-1) H'' -4B -2(1+u)C \bigr ]
\nonumber\\ & + & T^{(5)} \bigl [(1+u)G'' +(d-1)G' -B'-C \bigr
]\, .
\nonumber\eea
For clarity, we have omitted the indices of the tensors
$T^{(i)}$ defined in (\ref{3.7}). We also assume that the points
are separated so that the $\delta$-functions in (\ref{3.6}) can be
dropped.

The system of equations which determines the
graviton propagator is obtained by setting the scalar
coefficient of each independent tensor in (\ref{3.11}) to 0. We
focus first on the three equations
\bea
 C'(u) & = & - G''(u)  \label{3.12a} \\ B(u)+C(u) & = &
(1+u) G'(u) + (d-1) G(u) \label{3.12b} \\ -2B(u) & = &u(2+u)
G''(u) + (d-1) (1+u) G'(u)-2(d-1)G(u)\, . \label{3.12c}
\eea 
The first two equations can be integrated immediately to
give
\bea\label{3.13b}
     C(u) & = &- G'(u)  \\
     B(u) & = & (1+u)G'(u) +(d-1)G(u) 
\nonumber\eea
Integration constants are dropped so that all functions
vanish on the boundary. When (\ref{3.13b}) is substituted in
(\ref{3.12c}) we find an uncoupled differential equation for G(u),
namely
\be 0  = u(2+u) G''(u) + (d+1) (1+u) G'(u)
\label{3.14}
\ee
Comparing with (\ref{2.12}-\ref{2.13}) we see that this is exactly
the equation for a massless scalar propagator. This result
is consistent with the analysis of \cite{liutseytlin} in a
non-covariant gauge. The correctly normalized solution can be
obtained from older AdS literature \cite{prop} and is given by the
hypergeometric function
\bea 
\label{3.15}
 G(u) & = & \tilde C _d  (2u^{-1})^d F(d,{d+1 \over 2};d+1;-2u^{-1}) \\
\tilde C_d & = & {\Gamma ({d+1 \over 2}) \over (4\pi)^{(d+1)/2} d}
\nonumber\eea
An explicit solution in terms of elementary functions will be
given below.

We must now study the three remaining tensor equations in
(\ref{3.11}). After  eliminating $B(u)$ and $C(u)$ using (\ref{3.12a})
and (\ref{3.12b}) these become
\bea
0 & = &2G'' + 2(1+u) G' + 4(d-1) G + (d-1) H''  
\label{3.16a} \\ 0 & = &-u(2+u) H'' -2d(1+u) H' + 2d H  +4G
-2(1+u)G' -2(1+u) A   \label{3.16b} \\ 0 & = &(1+u) G' +2(d+1)
G - A' \label{3.16c} \eea
Since $G(u)$ is already known from (\ref{3.14}), this is an
over-determined system, and we must both obtain a solution and
check its consistency.

\medskip 
For this purpose we introduce the function $p(u)$
such that
\be p''(u)  = G(u) \qquad {\rm with} \qquad p(\infty) = p'
(\infty)  =0\, .
\label{3.17}
\ee In terms of this function, it is not hard to solve the
equations of (\ref{3.16a}-\ref{3.16c}) explicitly. From (\ref{3.16a}) and (\ref{3.16c}), and
the requirement that
$H(u)$ and $A(u)$ vanish at $u=\infty$, one obtains
respectively
\bea
 -(d-1) H(u) & = & 2p''(u) + 2(1+u) p'(u) + 4(d-2)
p(u) \label{3.18a} \\ A(u) & = &(1+u) p''(u) + (2d+1) p'(u) 
\label{3.18b} 
\eea
One can then check to see that (\ref{3.16b}) is also satisfied,
if we use (\ref{3.14}) and its integrated versions expressed in
terms of $p(u)$, which are given by
\bea\label{3.19}
&&u(2+u) p'''(u) +(d-1)(1+u) p''(u) -(d-1)p'(u)  = 0
\\ && u(2+u) p''(u) +(d-3) (1+u) p'(u) -2(d-2) p(u)  =0 
\nonumber\eea
A more useful expression for  $H(u)$ can be obtained by
eliminating $p(u)$ in (\ref{3.18a})  using (\ref{3.19}). This gives
\bea
 -(d-1) H(u) & = 2(1+u)^2 G(u) + 2(d-2) (1+u)
p'(u) & \label{3.20a}\\ -(d-1) p'(u) & = 2 \tilde C_d
(2u^{-1})^{d-1} F(d-1,{d+1 \over 2};d+1;-2u^{-1})\, . &
\label{3.20b}
\eea

Since $G(u)$ and $H(u)$ express the physical
content of the propagator and occur in integrals over
$AdS_{d+1}$ in applications to the AdS/CFT  correspondence,
explicit expressions in terms of elementary functions are
most useful. The first step to obtain them is to integrate
(\ref{3.14}) to obtain
\be G'(u) \sim 1/[u(2+u)]^{d+1 \over 2}\
\label{3.21}
\ee
The next integral can be evaluated, but the general form
of the result depends on whether $d$ is even or odd. For $d$
even we can rearrange the result given in \cite{grads} (see Sec. 2.263,(4))
to obtain
\be G(u) =  {\Gamma({d/2})(-)^{d/2}\over 4\pi^{d/2}}
\biggl\{ (1+u) \sum _{k=1}^{d/2}  { \Gamma(k- {1\over 2})\over
\pi^{{1\over 2}}\Gamma(k)}   {(-)^k \over [u(2+u)]^{k-{1\over 2}}} + 1
\biggr\}
\label{3.22}
\ee 
It can be verified directly that (\ref{3.22}) is the integral of
(\ref{3.21}) which vanishes at $u=\infty$. The normalization of
(\ref{3.22}) was chosen by requiring agreement with the
hypergeometric form (\ref{3.15}) as $u$ approaches 0. One can then
use (\ref{3.20a}) to obtain an analogous expression for $H(u)$,
namely
\bea\label{3.23}
&& H(u) =  -{\Gamma({d/2})(-)^{d/2}\over 4\pi^{d/2}}\times
\\
&& \biggl\{ {2(1+u)\over d-1}
 \sum _{k=1}^{d/2}  { \Gamma(k- {1\over 2})(-)^k \over
\pi^{{1\over 2}}\Gamma(k) (2k-3)}
\biggl[{2k-3\over [u(2+u)]^{k-{1\over 2}}} + {2k-d-1\over
[u(2+u)]^{k-{3\over2}}}\biggr] + 2(1+u)^2
\biggr\}\nonumber\eea These expressions may be worked out quite explicitly for
the special case of interest, $d=4$,
\bea
 G(u) &=&- {1 \over 8\pi ^2} \Bigl\{ {(1+u)
[2(1+u)^2 -3] \over  [u(2+u)]^{3\over2}}
          - 2\Bigr\} \label{3.24}\\ 
H(u) &=& {1 \over
12\pi^2}\Bigl\{ {(1+u) [6(1+u)^4 - 9(1+u)^2 +2] \over
[u(2+u)]^{3\over2}}
          - 6(1+u)^2\Bigr\}\ . \label{3.25}
\eea
For $d=2p+1$, the hypergeometric functions entering into
the definition of
$G(u)$ and $H(u)$ are degenerate, and reduce to elementary
functions involving logarithms. These expressions will not be
given here.

As in the case of the photon propagator it is worth  observing here that there 
is a more direct way of obtaining the gauge invariant functions $G(u)$
and $H(u)$ in which the compensating diffeomorphism (\ref{3.10}) and the
manipulation of a set of coupled equations for $G,H,A,B,C$ are
entirely avoided. The procedure is to substitute the ansatz (\ref{3.9}),
with gauge terms dropped, in the integral (\ref{3.5}). We then apply
the Ricci operator (\ref{3.6}) under the integral, using (\ref{3.11}) (with
diffeomorphism terms dropped). The various tensor terms are then
manipulated so as to isolate total derivatives ${\partial\over
\partial w^{\mu'}}$ which vanish by partial integration since
$D_{\mu'} T^{\mu'\nu'}(w)=0.$ There are 3 independent tensor
structures in the remaining terms, which directly give (\ref{3.14}) for
$G(u)$, and (\ref{3.16a}) and (\ref{3.16b}) for $H(u)$ (the latter with the
solution of (\ref{3.16c}) for $A(u)$ already in place).

Our final observation, from (\ref{3.15}) and (\ref{3.20a}), is
that $G(u)$ and $H(u)$ vanish on the boundary as $1/u^d$ and
$1/u^{(d-2)}$ respectively. One can then look back to
(\ref{3.12a}-\ref{3.12c}) and (\ref{3.18b}) and observe that the functions
$A(u),B(u)$ and $C(u)$ of the compensating diffeomorphism in
(\ref{3.10}) vanish as
$1/u^{(d-1)}, 1/u^d$ and $1/u^{(d+1)}$, respectively.  There
is then no boundary contribution in the integral of (\ref{3.6})
with a conserved stress tensor provided that $T_{00}(w)$
vanishes on the boundary. This is a consistency check on the
approach we have used. However, because the method is new, a
further check is also desirable. For this reason we show in
Section 5 that one obtains the same functions $G(u)$
and $H(u)$ from a more traditional approach. 
\space*{-.4in}\\

\section{The photon propagator in the Landau gauge}
\setcounter{equation}{0}
In this second part of the paper we  wish to obtain the photon and graviton propagators in $AdS_n$ ($n=d+1$) by a more 
standard covariant gauge--fixing procedure. We will employ techniques
developed in \cite{allenjacobs,allenturyn,turyn,mottola}, following quite 
closely \cite{mottola} where a simplified derivation of the graviton propagator
for 4--dimensional de Sitter space was given. 
In this and the following section
we will adopt the conventions introduced in \cite{allenjacobs}, which are 
followed throughout this earlier literature. 

\subsection{Notations}
Let $\mu(z,w)$ be the 
geodesic distance between $z$ and $w$. Then the basic tensors are the unit
tangents to the geodesics at $z$ and $w$:
\be \label{n}
n_{\mu}(z,w) = D_\mu \mu(z,w)  , \quad n_{\mu'}(z,w) = D_{\mu'} \mu(z,w)
\ee
together with the parallel transporter $g_\mu^{\,\nu'}(z,w)$ 
and the metric tensor 
itself at $z$ and $w$. We summarize in Table 1 some useful geometric
identities while Table 2 contains the dictionary that
translates between the two different sets of conventions used in the first and second part
of the paper. 
Note the the $AdS$ scale is always  set to 1, {\it i.e.} the curvature
scalar is
$R= -n(n-1)$.
\begin{center}
\renewcommand{\arraystretch}{2}
\begin{tabular}{||p{3in}||}
\hline
$g^{\;\;\nu'}_\mu n_{\nu'}=-n_\mu$\\
$D_\mu n_\nu=A\,(g_{\mu\nu}-n_\mu\,n_\nu)$\\
$D_\mu n_{\nu'}=C\,(g_{\mu\nu'}+n_\mu\,n_{\nu'})$\\
$D_\mu g_{\nu\rho'}=-(A+C)(g_{\mu\nu}n_{\rho'}+g_{\mu\rho'}n_\nu),$\\
where\\
$A(\mu)=\coth(\mu)$\\ 
$C(\mu)=-  {\displaystyle {1\over\sinh(\mu)}}$\\
with \\      
$A^2-C^2=1$\\ 
${\displaystyle{dA\over d\mu}=-C^2}$\\ 
${\displaystyle{dC\over d\mu}=-AC}$\vspace*{.1in} \\ 
\hline
\end{tabular}

\end{center}
\begin{center}{Table 1}\end{center}
\begin{center}
\renewcommand{\arraystretch}{2}
\begin{tabular}{||p{3in}||}
\hline
$n=d+1$\\
$u=\cosh(\mu)-1=2(x-1)$\\
${\displaystyle{n_\mu={\partial_\mu u\over \sqrt{u(u+2)}}}}$\\
${\displaystyle{g_{\mu\nu'}=\partial_\mu\partial_{\nu'} u + 
{\partial_\mu u \partial_{\nu'} u \over u+2}}}$\vspace*{.1in}\\ 
\hline

\end{tabular}

\end{center}
\begin{center}{Table 2}\end{center}
\subsection{Gauge choice and field equations for the photon propagator}

Let us write  the gauge potential as
\be
A_\mu = A_\mu^\perp + D_\mu \Lambda \, , \label{vectordecomp}
\ee
where
\be
 D^\mu A_\mu^{\perp} =0 \,.
\ee
A pleasant generic feature of  Anti de Sitter space (as opposed to de Sitter)
is the absence of normalizable zero modes. It follows that the 
decomposition (\ref{vectordecomp}) is unique, as long as one is careful
in demanding proper (fast enough)  falloff at the boundary of all the quantities involved. 
One has
\be
 A_\mu^\perp = P_\mu^{(1)\,\,\nu'} \,A_{\nu'}, 
\ee
where
\be
P_\mu^{(1)\,\,\nu'}  \equiv  g_{\mu}^{\,\nu'}\delta(z,w) -D_{\mu}\Box^{-1}D^{\nu'} \label{spin1proj}\,
\ee
is the projector onto transverse vectors. 

Let us now impose the Landau gauge
\be
 D^\mu A_\mu =0
\ee
which implies $\Lambda \equiv 0$. With this gauge choice the Maxwell equations read:
\be
-(g_{\mu \nu} \Box -R_{\mu \nu} )A^\nu = j_\mu \label{maxwell}
\ee
where the current is understood to be conserved, $D_\mu j^\mu =0$. The photon
propagator $G_{\mu \nu'}$ in the Landau gauge satisfies
\be
-( \Box +(n-1))G_{\mu \nu'}=  P_{\mu \nu'}^{(1)} \label{vectorpropequ}
\ee
 where we have used $R_{\mu \nu}=-(n-1)g_{\mu \nu}$ for $AdS_n$.
\subsection{Ansatz for the propagator using bitensors}
 We will now closely follow the manipulations in
Section III of \cite{allenjacobs}, where the spin 1 {\it massive} propagator was derived. 
The crucial difference between the discussion in \cite{allenjacobs} and our case is the
presence of the projector on the r.h.s. of (\ref{vectorpropequ}). We thus expect, and we are going to confirm, that the naive $m^2 \rightarrow 0$ limit
of the result in \cite{allenjacobs} does not give the correct Landau gauge massless
propagator. It gives instead a pure gauge part that can be added to
the massless propagator.

The most general ansatz for $G_{\mu \nu'}$ consistent with the
 symmetries of $AdS_n$ is 
 \be
G^{\;\;\nu'}_\mu=\alpha(\mu)g^{\;\;\nu'}_\mu+\beta(\mu)n^{\nu'}n_\mu \,.
\label{Gmunu'}
\ee
Let 
\be
\gamma = \alpha - \beta \,. \label{gamma}
\ee
Then using the identities in Table 1 (see \cite {allenjacobs}) one finds that the
gauge condition  $D^\mu G_{\mu \nu'}=0$ gives:
\be
-(n-1)C \alpha = \gamma' +(n-1)A \gamma \, ,\label{alphagamma'}\, .
\ee
We can apply the operator $-\left(\Box +(n-1)\right)$ onto
(\ref{Gmunu'}) to obtain the l.h.s. of (\ref{vectorpropequ})
\cite{allenjacobs}. We can extract an equation for $\gamma$ by
contracting the resulting bitensor with $n_\mu n_{\nu'}.$ The r.h.s. of
(\ref{vectorpropequ}) for $z\neq w$ is, after an integration by parts,
$D_\mu D_{\nu'} \Box^{-1}.$
The contraction with $n_\mu n_{\nu'}$ gives
\be
n^\mu P_{\mu\nu'}^{(1)} n^{\nu'}={d^2\over d\mu^2}\Box^{-1}.
\ee
Then we obtain the equation 
\be
\left({d^2\over d\mu^2}+(n+1)A(\mu){d\over d\mu}+ 2(n-1)\right)\gamma=-{d^2\over d\mu^2}\Delta_0\,, \label{gammaequ}
\ee
where $\Delta_0$ is the  massless scalar propagator\footnote{$\Delta_0$ is 
the same as $G(u)$ in equations (\ref{3.14},\ref{3.15}).} 
\be
\Box \Delta_0(\mu(z,w)) = -\delta(z,w) \label{Delta0}.
\ee
\subsection{Solving for $\gamma$}
Equation (\ref{gammaequ}) can be rewritten 
in terms of the $(n+2)$--dimensional Laplace operator:
\be
\left(\Box_{n+2}+ 2(n-1)\right)\gamma=-{d^2\over d\mu^2}\Delta_0\,. \label{gammaequlaplace}
\ee
Introducing the variable $x \equiv  \cosh^2 (\mu/2)=(u+2)/2$, the Laplace 
operator in $D$--dimensional $AdS$ is
\be
\Box_D=-x(1-x)\frac{ d^2}{{ d}x^2}- \frac{D}{2}(1-2x)\frac{ d}{{ d}x} \label{laplace}
\ee
and   (\ref{gammaequ}) reads, for $x \neq 1$:
\be
\left(-x(1-x)\frac{ d^2}{{ d}x^2}- \frac{n+2}{2}(1-2x)
\frac{ d}{{ d}x}+2(n-1) \right) \gamma = \frac{n-1}{2}(2x-1) \frac{ d}{{ d}x}\Delta_0 \label{gammaequx}
\ee
where we have used (\ref{Delta0}) to simplify the r.h.s\footnote{Our
$x$ is called $z$ in \cite{allenjacobs}--\cite{mottola}.}. Observe that the r.h.s. is an eigenvector of zero eigenvalue of the
differential operator on the l.h.s., up to a delta function term at $x=1$. This has a
simple physical interpretation: for $z \neq w$, the projector on the r.h.s of (\ref{vectorpropequ})
is a  pure gradient (see(\ref{spin1proj})), and thus a
 zero mode of the Maxwell operator.

To find a more explicit form of the r.h.s.  we use the following observation.
For any $f$,
\be
\frac{ d}{{ d}x}\left( \Box_D f \right)=\left( \Box_{D+2} + D \right)\frac{{ d}f}{{ d}x}\,.\label{observ}
\ee
It follows that, for $x \neq 1$,
\be
\left (\Box_{n+2}  +n \right) \frac{ d}{{ d}x}\Delta_0 =0 \,.\label{Delta0'}
\ee
It is now easy to solve for $\frac{ d}{{ d}x}\Delta_0$ by noting
that $\left(\Box_{n+2}  +n \right)$ is the  hypergeometric operator with parameters $a=n$, $b=1$, $c=n/2+1$. The solution with fast falloff at the boundary
($x \rightarrow \infty$) is, since  $a > b$ \cite{grads}
\be
(-x)^{-a}\,F(a,a+1-c;a+1-b;x^{-1}) =  (-x)^{-n/2}(1-x)^{-n/2}.
\ee
Normalizing the short distance ($x \rightarrow 1$) singularity as in
(\ref{Delta0}) we have
\be\label{delta0norm}
\frac{ d}{{ d}x}\Delta_0 = -{\Gamma\left({n\over
2}+1\right)\over 2^{n-1}n\pi^{n\over 2}}\;\frac{1}{[x(x-1)]^{\frac{n}{2}}}\,.
\ee

We now want to solve the inhomogeneous equation (\ref{gammaequx}). The
l.h.s. is $\left(\Box_{n+2}+ 2(n-1)\right)\gamma.$ By using the
relation (\ref{observ}) twice we find a massless operator acting on
$\int\int\gamma$:
\be
\Box_{n-2} \left(\int \int \gamma\right) = \int \int \left(\frac{n-1}{2}(2x-1) \frac{ d}{{ d}x}\Delta_0 \right) \label{gammaequint} \,,
\ee
or
\be
\left( -x(1-x)\frac{d}{dx} -\frac{n-2}{2}(1-2x) \right) \int \gamma =
{(n-1)\Gamma\left({n\over 2}-2\right)\over 2^{n+1}\pi^{n\over 2}}\;
{F(-n+4,1;-\frac{n}{2}+3; x)\over [x(x-1)]^{\frac{n}{2}-2}}
\ee
which is   a {\it first} order differential equation in $\int \gamma$ and can 
be solved by quadrature.

The two homogeneous
solutions of (\ref{gammaequx}) are readily obtained by recognizing
that the l.h.s. is 
 the hypergeometric operator with parameters $a=2$, $b=n-1$, $c=n/2+1$.
We then have for the general solution of (\ref{gammaequx}):
\be
\gamma  =  \gamma_{part} + a \,\gamma_1 + b\,\gamma_2, \label{gammageneral}
\ee
where
\bea
\gamma_{part} & = &{(n-1)\Gamma\left({n\over 2}-2\right)\over
2^{n+1}\pi^{n\over 2}}\;
\frac{d}{dx} \left[ 
\frac{\displaystyle \int_0^x dx'\:F(-n+4,1;- \frac{n}{2}+3;
x')}{\displaystyle [x(x-1)]^{\frac{n}{2}-1}}
  \right] \label{part}\\
\gamma_1 & = & \frac{2x-1}{[x(x-1)]^{\frac{n}{2}}} \\
\gamma_2 & =&  F(2,n-1;\frac{n}{2}+1,1-x) \label{gammageneralsol}\,,
\eea
For odd dimensions $n\geq 5$ the hypergeometric function in
(\ref{part}) is a polynomial of degree $n-4.$ For example, for $n=5$
\be
\gamma_{part}^{n=5}={1\over 32{\pi }^{2}}\,{\frac {2\,x-1}{\left
(x(x-1)\right )^{3\over 2}}}.
\ee
Notice that we have chosen a  basis of homogeneous solutions in which
$\gamma_1$ is singular at short distance 
($\sim 1/(x-1)^{n/2}$ as $x \rightarrow 1$) and has fast falloff at 
the boundary ($\sim 1/x^{n-1}$ for $x \rightarrow \infty$),
while $\gamma_2$ is smooth for $x \rightarrow 1$ and falls like $1/x^2$
at infinity. 

The constants $a$ and $b$ are determined by physical
considerations. First, we demand  the photon propagator
to have the same short distance singularity as in flat space. 
This requires
$\gamma \sim 1/(x-1)^{n/2-1}$ and hence $a \equiv 0$.
Second, we choose the boundary condition of fastest
possible falloff as $x \rightarrow \infty$, which fixes
$b \equiv (-1)^{\frac{n+1}{2}} {\Gamma\left({n+1\over 2}\right)\over 2\pi^{n-1\over 2}n(n-3)}$. With this choice, $\gamma \sim \log x/x^{n-1}$ as
$x \rightarrow \infty$. 
The final answer for the Landau gauge propagator (\ref{Gmunu'}) 
can be obtained by solving
(\ref{gamma}) and (\ref{alphagamma'}) for $\alpha$ and $\beta$:
\bea
\alpha(x) & = & {2\over n-1}x(x-1)\gamma'(x)+(2x-1)\gamma(x) \\
\beta(x) & = & {2\over n-1}x(x-1)\gamma'(x)+2(x-1)\gamma(x).
\eea
\subsection{Comparison to the new form of the propagator}
We now wish to make contact between this standard Landau gauge 
propagator
and the new form of the  propagator 
described in Section 2. To compare, we go back to the variable $u =2x-2$ 
and write our Landau gauge propagator in the same tensor 
basis used in (\ref{2.8}):
\be
G_{\mu \nu'}^{Landau} = - (\partial_\mu \partial_{\nu'} u){\tilde F}(u)
+ \partial_\mu \partial_{\nu'} {\tilde S}(u) \,.
\ee
Using the conversion formulas of Table 2, it is easy
to show:
\bea
-{\tilde F}(u) + {\tilde S}''(u)& =& -\alpha \\
{\tilde S}''(u) & = & \frac{\alpha}{u+2} + \frac{\beta}{u(u+2)}\,.
\eea

 One can show that if $\gamma$ satisfies
(\ref{gammaequx}) then ${\tilde F}$ obeys (\ref{2.12}). This is enough
to prove that ${\tilde F} = F$, since the solution of (\ref{2.12}) 
is uniquely specified (up to a factor) by the boundary condition of fastest falloff at
the boundary,  and the  
normalization is fixed in both cases by matching to the flat
space limit. As expected,  the ``Landau gauge'' $\tilde F$
coincides with  the ``universal'' $F$ found in Section 2. 

The  functional form of $\tilde S$ is instead specific to
 our Landau gauge 
choice. As a final consistency check,
one can  show that the ``gauge''
piece  $\partial_\mu \partial_{\nu'} {\tilde S}(u)$
gives no contribution when integrated with conserved currents.
Indeed,  $\tilde S(u) \sim 1/u^{n-3}$ as $u \rightarrow \infty$
and $\tilde S(u) \sim 1/u^{n/2-2}$ as $u \rightarrow 0$, 
which is enough to ensure the vanishing of the boundary
terms in the  integration by parts. 
It is interesting to observe that, for $u \neq 0$, the homogeneous solutions $\gamma_1$ and $\gamma_2$ do not
contribute to the physical term $\tilde F$. One may then be tempted to treat  $\gamma_1$ and $\gamma_2$ as pure
gauge artifacts and conclude that the constants $a$ and $b$ in (\ref{gammageneral}) can be arbitrarily specified. This is
however not quite correct. Changing $a$ and $b$ from the determined values would make $\tilde S$ too singular at the origin
or at infinity, so that  $\partial_\mu \partial_{\nu'} {\tilde S}(u)$  could not be ``gauged away'' from physical amplitudes.

\section{The graviton propagator in the Landau gauge}
\setcounter{equation}{0}
We now wish to obtain the graviton propagator in the Landau gauge. 
Our analysis will parallel in many aspects the discussion of the photon
in the previous section, although some new features will emerge. 

\subsection{Gauge choice and field equations for the graviton
propagator}

The linear fluctuations of the metric $h_{\mu \nu}$ can always be decomposed
as:
\be \label{tensordecomp}
h_{\mu \nu} =h_{\mu \nu}^{\perp}+D_{\mu}A_{\nu}^{\perp}+D_{\nu}A_{\mu}^{\perp}+
(D_\mu D_\nu -\frac{1}{n} g_{\mu \nu} \Box)B + \frac{1}{n}g_{\mu \nu} h
\ee 
where
\be
 D^\mu A_\mu^{\perp} = 
D^\mu h_{\mu \nu}^{\perp} = 
g^{\mu \nu }h_{\mu \nu}^{\perp}=0
\ee
and $h \equiv h_{\mu}^{\,\mu}$.
 The tensor
decomposition (\ref{tensordecomp}) can be uniquely inverted: 
\bea \label{tensorinversion}
B & = & \frac{n}{n-1}\left(\Box + \frac{R}{n-1} \right)^{-1} 
\Box^{-1} \left( D^{\mu'} D^{\nu'} -\frac{1}{n}g^{\mu' \nu'} \Box' \right)h_{\mu'\nu'} \\
A_{\mu}^{\perp}& =& -Q^{\,\nu'}_{\mu}D^{\nu'}h_{\mu'\nu'}	\\
h^{\perp}_{\mu \nu} &= & P_{\mu \nu}^{(2)\,\, \mu' \nu'}(z,w)\, h_{\mu' \nu'}\\
\label{spin2proj} P_{\mu \nu}^{(2)\,\, \mu' \nu'}(z,w) & \equiv & 
g_{(\mu}^{\, (\mu'} g_{\nu)}^{\, \nu')} \delta(z,w) -\frac{1}{n} g_{\mu \nu}
g^{\mu' \nu'}\delta(z,w) -2D_{(\mu}D^{(\mu'}Q_{\nu)}^{\,\nu')}\\
&& -\frac{n}{n-1}\left( D_{\mu} D_{\nu} -\frac{1}{n}g_{\mu \nu} \Box \right)
\left(\Box + \frac{R}{n-1} \right)^{-1} 
\Box^{-1} \left( D^{\mu'} D^{\nu'} -\frac{1}{n}g^{\mu' \nu'} \Box' \right)
\nonumber \eea
where   $(\cdot)$ indicates symmetrization with
strength 1. $Q^{\, \nu'}_{\mu}$  
obeys the equation
\be \label{Qmunu'}
  -(\Box +R/n )Q_{\mu}^{\;\; \nu'}= P^{(1)\;\nu'}_{\mu}\,,
\ee
where the right--hand side is the  transverse spin 1 projector (\ref{spin1proj}). Note that
this is {\it not} the same as (\ref{vectorpropequ}), the equation obeyed by 
the massless spin 1 propagator in the Landau
gauge, the difference being in the value of the ``mass''. 
It is understood the all the Green's functions in
(\ref{tensorinversion}--\ref{Qmunu'})  
obey the  boundary conditions appropriate for Anti de Sitter space, namely
fastest possible falloff at the boundary. Notice that in contrast to
the de Sitter case discussed in \cite{mottola} we do not have a contribution
to the r.h.s. of (\ref{Qmunu'}) from zero modes.

The covariant gauge choice which yields the simplest results is the ``Landau
gauge'':
\be
\label{landau}
D^{\mu} h_{\mu \nu} = \frac{1}{n} D^{\nu} h \, . 
\ee
With this gauge choice, $B= A_{\mu}^{\perp} \equiv 0$ and
the graviton propagator  can be written as:
\be \label{fullprop}
G_{\mu \nu \mu' \nu'}(z,w) 
 = g_{\mu \nu} g_{\mu' \nu'} T(\mu(z,w))+G^{(2)}_{\mu \nu \mu' \nu'}(z,w)
\ee
with $G^{(2)}_{\mu \nu \mu' \nu'}(z,w)$ transverse and traceless:
\bea
D^{\mu} G^{(2)}_{\mu \nu \mu' \nu'}&=&D^{\mu'} G^{(2)}_{\mu \nu \mu' \nu'}= 0 \label{trans}\\
g^{\mu \nu}G^{(2)}_{\mu \nu \mu' \nu'}&=&g^{\mu' \nu'}G^{(2)}_{\mu \nu \mu' \nu'}=0
\label{notrace}
\,.
\eea

Projection of the linearized Einstein equations onto the ``pure trace'' and 
``transverse symmetric traceless'' subspaces gives:
\bea
\label{Sequation}
\left( \Box + \frac{R}{n-1} \right) T &=& \frac{\delta(z,w)}{(n-2)(n-1)} \\
\label{G2equation} \left(- \Box + \frac{2R}{n(n-1)} \right) G^{(2)}_{\mu \nu \mu' \nu'}& =& 
 P_{\mu \nu\mu' \nu'}^{(2)} \label{G2equ}
\eea
where the spin 2 projector has been defined in (\ref{spin2proj}).
We now proceed to solve (\ref{Sequation}) and (\ref{G2equation}).

\subsection{Solving for $T$}

Using $R=-n(n-1)$ we see that (\ref{Sequation}) defines a scalar propagator
of $m^2= n$.
 Introducing the variable $x \equiv  \cosh^2 (\mu/2)=(u+2)/2$ 
 we recognize (\ref{Sequation}) (see (\ref{laplace})) as the hypergeometric
equation of parameters $a=n $, $b=-1 $, $c=n/2 $. The fastest falloff solution
with  short distance 
singularity normalized as in  (\ref{Sequation}) is  
\cite{prop} \cite{grads}:
\be \label{Thyper}
T(x) = -\frac{1}{(n-2)(n-1)}\frac {\Gamma (n)\Gamma \left({n\over2}+1\right)}{\Gamma (n+2){\pi }^{n\over 2}{2
}^{n}}\;
\frac{1}{x^n}F(n,\frac{n}{2}+1;n+2;\frac{1}{x})\,.
\ee
Observe that using (\ref{observ}) the operator $\left( \Box_n -n \right)$ on $T$ changes to
 $\Box_{n+2}$ on $\frac{d}{dx} T$, so that $T(x)$ could also
 be obtained by quadrature.
$T(x)$ is actually an algebraic function for odd $n$. For example, for
$n=5$:
\be
T^{n=5}(x) ={\frac {1}{768 \pi^2}}\,{\frac {128\,{x}^{4}-256\,{x}^{3}+144\,{x}^{2}-16\,x
-1}{\left (x-1\right )^{3\over2}{x}^{3\over2}}}-{\frac {2\,x-1}{{12
\pi }^{2}}}\,.
\ee

\subsection{Ansatz for  $G^{(2)}_{\mu\nu\mu'\nu'}$ in terms of bitensors}

We want to write  the most general ansatz for  $G^{(2)}_{\mu\nu\mu'\nu'}$ consistent
with $AdS$ symmetry.
The 5 independent fourth rank bitensors which
are symmetric under $\mu \leftrightarrow \nu$, $\mu' \leftrightarrow \nu'$
and  $(\mu, \nu) \leftrightarrow (\mu', \nu')$ are taken to be:
\bea \label{O's}
{\cal O}_{\mu \nu  \mu' \nu'}^{(1)} &=& g_{\mu \nu}g_{\mu' \nu'} \nonumber \\
{\cal O}_{\mu \nu  \mu' \nu'}^{(2)} &=& n_\mu n_\nu n_{\mu'} n_{\nu'} \nonumber \\
{\cal O}_{\mu \nu  \mu' \nu'}^{(3)} &=& g_{\mu \mu'} g_{\nu \nu'}+
                                  g_{\mu \nu'} g_{\nu \mu'}\\
{\cal O}_{\mu \nu  \mu' \nu'}^{(4)} &=& g_{\mu \nu} n_{\mu'} n_{\nu'}+ 
g_{\mu' \nu'} n_{\mu} n_{\nu} \nonumber \\
{\cal O}_{\mu \nu  \mu' \nu'}^{(5)} &=& g_{\mu \mu'} n_{\nu} n_{\nu'}+g_{\mu \nu'} n_{\nu} n_{\mu'} + g_{\nu \nu'} n_{\mu} n_{\mu'} +g_{\nu \mu'} n_{\mu} n_{\nu'} \,.
\nonumber
\eea
 This basis is linearly related that of (\ref{3.7}),  as shown in  Table 3. We then have
\be
G^{(2)}_{\mu\nu\mu'\nu'} = \sum_{i=1}^5 \,G_i(\mu)\,{\cal O}_{\mu \nu  \mu' \nu'}^{(i)}\,.
\label{5tensoransatz}
\ee
The tracelessness conditions (\ref{notrace}) imply:
\bea
 G_2 &=&4G_5-nG_4 \label{trace1}\\
 G_1 &=& -{1\over n}(2G_3+G_4)\,.
\eea
Define $f$ and $g$ as
\bea
f & \equiv &G_5-G_3\\
  g & \equiv & (n-1)G_4-2G_3\,.
\eea 
Then the transversality conditions (\ref{trans}) imply:
\bea
&& f'(\mu)+nA(\mu)f(\mu)+{n\over 2}C(\mu)g(\mu) - \left({n(n-1)\over 2}-1\right)C(\mu)G_4(\mu)=0\\
&& g'({\mu})+nA(\mu)g(\mu)+2nC(\mu)f(\mu)=0 \label{transv2}\,.
\eea
It is easy to see that the five functions $G_{i}$ can be determined immediately 
from the single function
$g$.  To find $g$, we need to use the equation of motion (\ref{G2equ}).
Operating  $\left(- \Box -2  \right)$ onto the ansatz (\ref{5tensoransatz})
(see formulas (A4) in \cite{turyn}) we get the l.h.s. of  (\ref{G2equ}). Contracting the resulting bitensor 
with $n_\mu n_\nu n_{\mu'} n_{\nu'}$  we can extract an expression containing
only $g$. Equ.(\ref{G2equ}) then implies:
\be
-\left({n-1\over n}\right)\:\left( x(1-x)\frac{d^2}{dx^2} +({n\over 2}+2)(1-2x) \frac{d}{dx} -2(n+1) \right)g=
n^{\mu} n^{\nu} n^{\mu'} n^{\nu'} P_{\mu \nu\mu' \nu'}^{(2)}\,. \label{gequ}
\ee

\subsection{Finding the r.h.s.}

We now need a more explicit form for the spin 2 projector (\ref{spin2proj}), which
appears in (\ref{gequ}).
Define $\Delta_0$ as in (\ref{Delta0}) and 
\be\label{Delta1}
\left (\Box-n \right)\Delta_1 (z,w) = -\delta(z,w)\,.
\ee
Further, as in the case of the photon propagator (see (\ref{Gmunu'})),
let us write the  ansatz for $Q^{\;\;\nu'}_\mu$ in (\ref{Qmunu'}) as
\be\label{expq}
Q^{\;\;\nu'}_\mu=\tilde \alpha(\mu)g^{\;\;\nu'}_\mu+\tilde \beta(\mu)n^{\nu'}n_\mu.
\ee
Going through the same steps as in Section 4.3, we define
\be
\tilde \gamma=\tilde \alpha-\tilde \beta.
\ee  
and we arrive at 
\be
\left({d^2\over d\mu^2}+(n+1)A(\mu){d\over d\mu}\right)\tilde \gamma=
-{d^2\over d\mu^2}\Delta_0 \,.\label{gammatilde}
\ee
Using (\ref{Delta0}) we can rewrite  (\ref{gammatilde}) as
\be
\left({d^2\over d\mu^2}+(n+1)A(\mu){d\over d\mu}\right)\left[\tilde \gamma-{n-1\over 2}\Delta_0\right]=0\,. \label{gammatildebis}
\ee
We require that $\tilde \gamma$ obeys the usual boundary condition of fastest
falloff at infinity and matches at short distance the flat space singularity.
Since the homogeneous solutions of (\ref{gammatildebis}) are either too singular at short
distance or fall off too slowly at infinity, we choose the particular solution:
\be\label{gamma7}
\tilde \gamma ={n-1\over 2}\Delta_0 \,.
\ee

We are now in the position to evaluate the r.h.s. of the basic equation (\ref{gequ}).
Let $z\neq w.$ From the expansion (\ref{expq}) of $Q$ we find
\be
n^\mu n^\nu n_{\mu'} n_{\nu'} \left[ -2 D_{(\mu} D^{\mu'} Q_{\mu)}^{\nu'} \right] = 2 {d^2\over d\mu^2} \gamma
\ee
To evaluate the contribution of the last term in $P^{(2)}$ in (\ref{spin2proj}), note that
\bea
&&\left( \Box +{R\over n-1}\right)^{-1}\,\Box^{-1}={ n-1\over R}\,\left(\Delta_{0}-\Delta_1\right)\\
&&n^\mu n^\nu \left( D_\mu D_\nu-{1\over n} g_{\mu\nu}\Box\right)\,=\,{d^2\over d\mu^2}-{1\over n}\: \Box \nonumber
\eea
Then we find
\be\label{rhs}
n^\mu n^\nu n^{\mu'} n^{\nu'} P^{(2)}_{\mu\nu\mu'\nu'}=2 {d^2\over d\mu^2} \gamma-{n\over R}\left[ {d^4\over d\mu^4}(\Delta_0-\Delta_1)+{2\over n}
\,{d^2\over d\mu^2}\Box\Delta_1-{1\over n^2}\, {d^2\over d\mu^2}\Box^2\Delta_1\right]
\ee
Now recalling (\ref{gamma7}), using the equations obeyed by $\Delta_0$
and $\Delta_1$ and changing to the variable $x$, $\mu={\rm arccosh}(2x-1)$, we can rewrite (\ref{rhs}) as
\be
n^\mu n^\nu n^{\mu'} n^{\nu'} P^{(2)}_{\mu\nu\mu'\nu'}=S_0(x)+S_1(x),
\ee
where $S_0(x)$ and $S_1(x)$ are defined as
\bea
S_0(x)&=&{1\over 8}\,{\frac {n (n+1)\left (2\,x
-1\right )}{x\left (x-1\right )}}\;{\frac {d}{dx}}{\Delta_0}(x)\label{s0}\\
S_1(x)&=&{1\over 4}\,{(n+1)(4(n-1)\:x(x-1)+n)\over n}\;{\frac {d^2}{dx^2}}{\Delta_1}(x)\label{s1}\,.
\eea
The expression for ${\frac {d}{dx}}{\Delta_0}$ was given in (\ref{delta0norm}).
To find an explicit functional form for ${\frac {d^2}{dx^2}}{\Delta_1}$, we use (\ref{observ}) twice to get, for $x \neq 1$:
\be
\left( \Box_{n+4} +n+2 \right) {\frac {d^2}{dx^2}}{\Delta_1}(x) =0
\ee
which is the same as (\ref{Delta0'}) with $n \rightarrow n+2$. The fast falloff solution normalized as in (\ref{Delta1}) is
\be
{\frac {d^{2}}{d{x}^{2}}}{\Delta_1}(x)={\frac {\Gamma \left({n\over 2}+1\right)}{{2
}^{n}{\pi }^{n\over 2}}}{1\over\left (x\left (x-1\right )\right )^{{n\over 2}+1}}.
\ee
We thus get
\bea
S_0(x)&=& -{(n+1)\Gamma\left({n\over
2}+1\right)\over 2^{n+2}\pi^{n\over 2}}\;\frac{2x-1}{(x(x-1))^{ \frac{n}{2}+1}}\,\\
S_1(x)&=& {(n+1)\,\Gamma\left({n\over
2}+1\right)\over 2^{n+2}n\pi^{n\over 2}}\;\frac{4(n-1)\:x(x-1)+n}{(x(x-1))^{ \frac{n}{2}+1}}.
\eea
Observe that $S_0$ and $S_1$ have opposite symmetry properties under ``antipodal reflection'' $x\rightarrow 1-x.$

\subsection{Solving for $g$}
If we imagine to turn off the source $S_1$, the equation (\ref{gequ}) is the same as the corresponding photon equation (\ref{gammaequx}) with 
$n\rightarrow n+2$ and a different normalization of the source. The functional form of a particular solution $g_{part}^0$ corresponding to the source $S_0$ can  thus be read from (\ref{part}).
Accounting for  the 
different normalization we find:
\be \label{gpart}
g_{part}^0(x)  ={n(n+1)\Gamma\left({n\over 2}-1\right)\over
(n-1)2^{n+2}\pi^{n\over 2}}\;
\frac{d}{dx} \left[ 
\frac{\displaystyle \int_0^x dx'\:F(-n+2,1;- \frac{n}{2}+2;
x')}{\displaystyle [x(x-1)]^{\frac{n}{2}}}
  \right]
\ee
For odd $n$, the hypergeometric function in (\ref{gpart}) is actually a polynomial of degree $n-2$.

As in the case of the photon, for $x\neq 1,$ $S_0$ is an eigenfunction
of the  differential operator on the l.h.s. of (\ref{gequ}) with zero eigenvalue. It turns out that the source $S_1$ is also an eigenfunction of the same operator with eigenvalue $-{(n-2)(n-1)\over n}.$ Thus the particular solution $g_{part}^1$ is
\be
g_{part}^1=-{n\over (n-2)(n-1)}\;S_1.
\ee
The general solution of (\ref{gequ}) is then  given by (compare with (\ref{part}--\ref{gammageneralsol})): 
\bea
g &=& g_{part}^1 +g_{part}^0 + \tilde a g_1 + \tilde b g_2 \\
g_1 & = & \frac{2x-1}{[x(x-1)]^{\frac{n}{2}+1}} \\
g_2 & =&  F(2,n+1;\frac{n}{2}+2,1-x) \,.
\eea
Notice that $g_1$ is singular at short distance ($\sim 1/(x-1)^{n/2+1}$ as $x \rightarrow 1$) and falls like $1/x^{n+1}$ as $x \rightarrow \infty$,
while $g_2$ is smooth at $x=1$ and falls like $1/x^2$ at infinity.
The constant $\tilde a$ is determined by demanding  that the graviton propagator has the same short distance singularity as in flat space,
which requires $g \sim /(x-1)^{n/2-1}$. This fixes $\tilde a \equiv \frac{n(n+1)\Gamma({n\over2}+1)}{(n-2)(n-1)2^{n+2} \pi^{n\over2}}$. The requirement of fastest possible falloff at the boundary
determines $\tilde b \equiv (-1)^{\frac{n-1}{2}}\frac{\Gamma(\frac{n+3}{2}) n}{(n+2)(n-1)^2 \pi^{\frac{n-1}{2}}}$.  With this choice,
 $g \sim \log x/x^{n+1}$ as $x \rightarrow \infty$. 

For example, in five dimensions we have
\be
g^{n=5}=-\frac{15}{1024 \,\pi^2}\;
\frac{8\,{x}^{3}-4\,{x}^{2}-6\,x+5}{x^{\frac{7}{2}} \, (x-1)^{\frac{3}{2}}} +\frac{15}{56 \, \pi^2}\,F(2,6;\frac{9}{2};1-x)\,.
\ee

The final answer for the transverse traceless part of the graviton propagator is given by (\ref{5tensoransatz}), where the functions 
$G_i$ are obtained by
solving (\ref{trace1}--\ref{transv2}) in terms of $g$:
\bea
G_1(x)& =& {1\over n(n-2)}\,\left({4(x-1)^2x^2\over n+1}\;g''(x)+\right.\nonumber\\
&&4x(x-1)(2x-1)\;g'(x)+(4nx(x-1)+n-2)g(x)\Bigg)\\
G_2(x)& =& -4(x-1)^2\left({x^2\over n(n+1)}\;g''(x)+{2(n+2)x\over n(n+1)}\;g'(x)+g(x)\right)\\
G_3(x)& =& -{1\over 2(n-2)}\,\left({4(n-1)x^2(x-1)^2\over n(n+1)}g''(x)+ \nonumber \right.\\
&&{4(n-1)x(x-1)(2x-1)\over n}\;g'(x)+(4(n-1)x(x-1)+n-2)\;g(x)\Bigg)\\
G_4(x)& =& -{4x(x-1)\over n(n-2)}\left({x(x-1)\over n+1}\;g''(x)+(2x-1)\;g'(x)+n\;g(x)\right)\\
G_5(x)& =&-{x-1\over n-2}\left({2(n-1)x^2(x-1)\over n(n+1)}\;g''(x) + \right. \nonumber \\
&& {x(4(n-1)x-3n+4)\over n}\;g'(x)+(2(n-1)x-n+2)\;g(x)\Bigg)                      \,.
\eea

The full graviton propagator in the Landau gauge is given by (\ref{fullprop}), so that its components in
the tensor basis (\ref{O's}) are simply ($T+G_1$, $G_2$, $G_3$, $G_4$, $G_5$), where the trace term $T$ was determined in  (\ref{Thyper}).

\subsection{Comparison to the new form of the propagator}

To make contact with the discussion in Section 3, we go back to the variable $u = 2x-2$ and write the Landau gauge graviton
propagator in a tensor basis analogous to (\ref{3.9}):
\bea
G^{Landau}_{\mu \nu \mu' \nu'}& =& (\partial_\mu \partial_{\mu'}
u\,\partial_\nu \partial_{\nu'} u+\partial_\mu \partial_{\nu'}
u\,\partial_\nu \partial_{\mu'} u) \,\tilde G(u) +g_{\mu \nu} g_{\mu'
\nu'} \,\tilde H(u) \\
&& +\,\partial_{(\mu} [\partial_{\nu)} \partial_{\mu'} u
\,\partial_{\nu'}u \,\tilde X(u)]+ \partial_{(\mu'} [\partial_{\nu')} \partial_{\mu} u \,
\partial_{\nu}u \,\tilde X(u)] \nonumber \\
&&  +\,\partial_{(\mu} [\partial_{\nu)}u \,\partial_{\mu'} u
\,\partial_{\nu'}u \,\tilde Y(u)] +\partial_{(\mu'} [\partial_{\nu')}u
\,\partial_{\mu} u \,\partial_{\nu}u \,\tilde Y(u)] \nonumber \\
&& + \,\partial_\mu [\partial_\nu u\,\tilde Z(u)] \,g_{\mu' \nu'} +
\partial_{\mu'} [\partial_{\nu'}u \,\tilde Z(u)] \,g_{\mu \nu} \,.\nonumber
\eea

\renewcommand{\arraystretch}{1.2}
\begin{center}
\begin{tabular}{|lrrrrr|}
\hline
$T_1=$&${\cal O}_1$&&&&\\
$T_2=$&&$[u(u+2)]^2{\cal O}_2$&&&\\
$T_3=$&&$2u^2{\cal O}_2$&$+{\cal O}_3$&&$-u{\cal O}_5$\\
$T_4=$&&&&$u(u+2){\cal O}_4$&\\
$T_5=$&&$4u^2(u+2){\cal O}_2$&&&$-u(u+2){\cal O}_5$\\
\hline
\end{tabular}\\ \medskip
Table 3
\end{center}


Using the conversion formulas between the two tensor basis (\ref{3.7}) and (\ref{O's}) given in Table 3, one can show:
\bea
\tilde Y'(u) & =& \frac{1}{4} \frac{G_2(u)+2u^2G_3(u)+4uG_5(u)}{u^2(u+2)^2}\\
  \tilde X'(u) & =& -\frac{G_5(u)+uG_3(u)}{u(u+2)}-2\tilde Y(u)\\
\tilde Z'(u) & = &\frac{G_4(u)}{u(u+2)}-2\tilde X(u) -2(1+u)\tilde Y(u) \\
\tilde G(u) & = & G_3(u) -2 \tilde X(u)\\
\tilde H(u) & = & T(u)+G_1(u)-2(1+u) \tilde Z(u)\,.
\eea

As in the case of the photon, one expects  the ``Landau gauge''
$\tilde H$ and $\tilde G$ to be exactly the ``universal'' $H$ and $G$ found 
in Section 3. We directly checked this  for $n=5$ working with the explicit expressions 
for $G_i$ and $T$. It is worth commenting on some interesting features of this check. 
First, the homogeneous solutions $g_1$ and $g_2$ do 
not contribute at all to $\tilde H$ and $\tilde G$, at least for $u \neq 0$. They however do contribute to 
$\tilde X$, $\tilde Y$, $\tilde Z$.  The  role of the homogeneous solutions is again  to ensure the proper falloff and short distance singularity of the
``gauge terms'', so that they can be removed when integrating the propagator with conserved currents.
 Second, the contribution of $g^1_{part}$ to the physical functions $\tilde H$, $\tilde G$ is precisely removed by  adding 
 the trace term $T$. Therefore there is no effective propagation
of the scalar of $m^2 = n$.

\newpage
\section*{Acknowledgments}
It is a pleasure to thank Richard Brower and  Emil Mottola for useful discussions.

The research of E.D'H is supported in part by NSF Grants
No. PHY-95-31023 and PHY-97-22072,
D.Z.F.  by
NSF Grant No. PHY-97-22072, S.D.M., A.M. and L.R. by D.O.E. cooperative agreement
DE-FC02-94ER40818. L.R. is  supported in part by INFN `Bruno Rossi' Fellowship

\end{document}